\documentclass[letter,twocolumn]{jpsj3}
\usepackage{txfonts}
\usepackage{subfigure}
\usepackage{bm}
\usepackage{xcolor}
\title{Theory of Defect-Induced Kondo Effect in Graphene:
\\Numerical Renormalization Group Study}

\author{\name{Taro \surname{Kanao}}\thanks{E-mail address: kanao@hosi.phys.s.u-tokyo.ac.jp}, \name{Hiroyasu \surname{Matsuura}}, and \name{Masao \surname{Ogata}} 
}
\inst{\address{Department of Physics, the University of Tokyo, Bunkyo, Tokyo 113-003, Japan} 
}
\abst{An effective model that describes the Kondo effect due to a point defect in graphene is developed, taking account of the electronic state and the lattice structure of the defect. 
It is shown that this model can be transformed into a single-channel pseudogap Anderson model with a finite chemical potential. 
On the basis of the numerical renormalization group method, it is clarified that the experimentally observed gate-voltage dependence of the Kondo temperature is understood in this framework.}

\kword{graphene, Kondo effect, point defect, pseudogap Anderson model, numerical renormalization group method}

\begin{document}
\maketitle

Recently, the Kondo effect in graphene has attracted much attention~\cite{Sengupta2008,Vojta2010,Zhu2010,Haase2011,Kotov2010} because exotic Kondo effects due to the characteristic nature of Dirac electrons in graphene~\cite{CastroNeto2009} are expected. 
It is reported that the Kondo effect is induced by point defects in graphene~\cite{Chen2011}. 
In this experiment, a negative magnetoresistance is observed, which indicates that the Kondo effect is the magnetic origin. 
Since the Kondo effect in usual metals is caused by magnetic impurities, this experiment suggests that the defects in graphene behave as magnetic impurities. 
The magnetic behavior of the defects in graphene is also suggested by recent magnetization mesurement of ion-irradiated graphene~\cite{Nair2012}. 
Although there is no spin at the point defects in the usual metals, the magnetic behavior of the defects is known in semiconductors, particularly silicon~\cite{Watkins1963}. 
In graphene, defects are introduced by ion irradiation. 
The most probable defects after irradiation are single vacancies, as shown by a scanning tunneling microscopy (STM) study~\cite{Ugeda2010}. 
The electronic states of point defects in graphene have been studied by density functional theory (DFT) calculations\cite{Lehtinen2004,Yazyev2007}, and the appearance of localized spin has been suggested, as in the case of the point defect in silicon. 

Although there have been several theoretical studies on the Kondo effect in graphene~\cite{Sengupta2008,Vojta2010,Zhu2010,Haase2011,Kotov2010}, only the adatom cases~\cite{Sengupta2008, Vojta2010, Zhu2010, Kotov2010} and the resonant scatterer case~\cite{Haase2011} have been discussed. 
No study has focused on the localized spin in the point defect itself as a cause of the Kondo effect. 
In addition, experimentally, the Kondo temperature depends on the gate voltage~\cite{Chen2011}. 
In particular, around the charge-neutrality point, the Kondo temperature shows a nonzero minimum value and almost symmetric dependence for the positive and negative gate voltages. 
These behaviors have not been understood theoretically. 

In this paper, in order to understand the properties of the Kondo effect and the related phenomena, we develop an effective model in which we take account of the electronic state and the lattice structure around the defect. 
We then analyze the model by the numerical renormalization group method~\cite{Wilson1975}. 
The relevance of the results to the experimental results is discussed. 
We show that the experimentally observed gate-voltage dependence of the Kondo temperature is understood in this framework. 

\begin{figure}
	\centering
	\includegraphics[width=4cm]{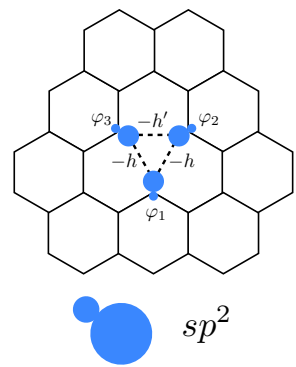}
	\caption{(Color online) Schematic picture of a point defect (single vacancy) in graphene and $sp^2$ orbitals (circles) around it ($\varphi_i$, $i=1,2$, and $3$). 
	$-h$ and $-h'$ are transfer integrals between them. }
	\label{fig_defect}
\end{figure}

In the case of silicon, the electronic states and the lattice structures of point defects have been studied in detail~\cite{Watkins1963}. 
It has been known that the electronic state around a point defect can be understood by considering molecular orbitals that consist of four $sp^3$ orbitals around the defect. 
Here, the same treatment is applied to the point defect in graphene. 

Figure \ref{fig_defect} shows the schematic picture of the point defect (single vacancy) in graphene. 
Around the point defect, there are three $sp^2$ orbitals, $\varphi_i$, (circles) which do not form covalent bond, where $i$ indicates one of the three sites ($i=1$, $2$, or $3$) around the defect. 
There are also $\pi$ orbitals on carbon atoms which are not shown in Fig. \ref{fig_defect}. 
The hybridization between the $sp^2$ and $\pi$ orbitals depends on the lattice structure of the defect. 
From STM~\cite{Ugeda2010} and transmission electron microscope experiments~\cite{Meyer2008} and DFT calculations~\cite{El-Barbary2003}, two possible lattice structures of the point defect are suggested: one is a threefold symmetric structure that preserves the original symmetry, and the other is a twofold symmetric structure in which one of the C-C distances (between sites $2$ and $3$ in Fig. \ref{fig_defect}) is shortened. 
While the former has a planar structure, the latter has out-of-plane displacement. 
When the defect has the planar structure, there is no hybridization between the $sp^2$ and $\pi$ orbitals. 
However, with out-of-plane displacements, finite hybridization between the $sp^2$ orbital, $\varphi_i$, and the two $\pi$ orbitals at the other sites, $j\neq i$, appears. 
From the DFT calculations~\cite{El-Barbary2003}, it is suggested that the twofold symmetric structure is more stable owing to the Jahn-Teller effect. 

First, we consider an isolated cluster that consists of three $sp^2$ orbitals, as depicted in Fig. \ref{fig_defect}, as a model for the localized state of $sp^2$ orbitals around the point defect. 
The tight-binding approximation is applied to this cluster. 
In Fig. \ref{fig_defect}, $-h$ and $-h'$ are transfer integrals between the $sp^2$ orbitals ($h,h'>0$). 
The threefold symmetric structure corresponds to the case with $h=h'$, and the twofold symmetric one to $h'>h$, assuming that the distance between sites $2$ and $3$ is shorter than the others. 

The energies and the wave functions of this cluster model are
\begin{eqnarray}
	&\begin{array}{cc}
		E_3=h',&\psi_3=\frac{1}{\sqrt{2}}(\varphi_2-\varphi_3),\\
		E_2=\frac{-h'+\sqrt{8h^2+h'^2}}{2},&\psi_2=C_+\varphi_1-\frac{1}{\sqrt{2}}C_-(\varphi_2+\varphi_3),\\
		E_1=\frac{-h'-\sqrt{8h^2+h'^2}}{2},&\psi_1=C_-\varphi_1+\frac{1}{\sqrt{2}}C_+(\varphi_2+\varphi_3),
	\end{array}&\label{eq_energy_wave_function}
\end{eqnarray}
with $C_\pm=\left[\frac{1}{2}\left(1\pm\frac{h'}{\sqrt{8h^2+h'^2}}\right)\right]^{1/2}$.
There are three electrons on these $sp^2$ orbitals. 
Since the number of electrons is odd, there is an unfilled level and nonzero spin appears. 
Interpreting the Kohn-Sham eigenvalues of the DFT calculations~\cite{El-Barbary2003} in terms of these tight-binding energy levels, $h'/h$ is semiquantitatively estimated to be $h'/h\simeq5$. 
Then, $C_\pm$ becomes $C_+\simeq1,C_-\ll 1$, and the wave function of the lowest level $\psi_1$ becomes the bonding state, $\psi_1=(\varphi_2+\varphi_3)/\sqrt{2}$, 
while the second state becomes $\psi_2=\varphi_1$. 
Thus $\varphi_1$ becomes the highest occupied molecular orbital of this cluster. 
It should be noted that in a divacancy case, four electrons are present in the $sp^2$ orbitals and the localized spin does not appear. 

\begin{figure}
	\centering
		\includegraphics[width=4cm]{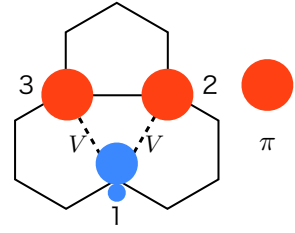}
	\caption{(Color online) Schematic picture of the electronic state around the defect. 
	Two of the three $sp^2$ orbitals ($i=2$, $3$) form a covalent bond. 
	The object at site $1$ shows the active $sp^2$ orbital and the circles at sites 2 and 3 the $\pi$ orbitals. 
	$V$ is the amplitude of the hybridization between the active $sp^2$ orbital and the two $\pi$ orbitals. }
	\label{fig_distortion}
\end{figure}

In the following we consider the structure shown in Fig. \ref{fig_distortion}, where only one $sp^2$ orbital ($i=1$) is active and it hybridizes with the $\pi$ orbitals at sites $2$ and $3$. 
(The bonding state of $sp^2$ orbitals at sites $2$ and $3$ is neglected.)
We assume the hybridization term is given by 
\begin{eqnarray}
	H_{\rm{hyb}}=\sum_{\sigma=\uparrow\downarrow}\left[V\left(a^\dagger_{2\sigma}+a^\dagger_{3\sigma}\right)d_{\sigma}+\mathrm{h. c. }\right]\label{eq_hybridization},   
\end{eqnarray}
where $a_{i\sigma}$ is an annihilation operator of a $\pi$ electron at site $i$ with spin $\sigma=\uparrow,\downarrow$ and $d_\sigma$ is an annihilation operator of an electron on the $sp^2$ orbital at site $i=1$. 
$V$ is the amplitude of hybridization. 
Taking account of the localized nature of the active $sp^2$ orbital in the defect, we assume that the effective Hamiltonian for this orbital becomes 
\begin{eqnarray}
	H_{\rm{def}}=\sum_{\sigma}(\epsilon_{sp^2}-\mu)d^\dagger_{\sigma}d_{\sigma}+Ud^\dagger_{\uparrow}d_{\uparrow}d^\dagger_{\downarrow}d_{\downarrow},\label{eq_defect_hamiltonian}
\end{eqnarray}
where $\epsilon_{sp^2}$, $\mu$, and $U$ are the energy level, chemical potential, and Coulomb interaction on the $sp^2$ orbital, respectively. 
The Coulomb interaction term is essential for describing the Kondo effect. 

The conduction electron states are formed by $\pi$ orbitals in graphene. 
Ignoring the Coulomb interaction in the $\pi$ orbitals, the conduction electron states are described by the following tight-binding model with a nearest-neighbour transfer integral $t(>0)$~\cite{CastroNeto2009}, 
\begin{eqnarray}
	H_{\rm{gra}}=-t\sum_{\langle ij\rangle\sigma}\left(a^\dagger_{i\sigma}b_{j\sigma}+\rm{h.c.}\right)-\mu N,\label{eq_graphene_hamiltonian}
\end{eqnarray}
where $b_{j\sigma}$ is an annihilation operator of a $\pi$ electron at site $j$ of the {\it B} sublattice, and $N$ is the total number of $\pi$ electrons. 
In eq. (\ref{eq_hybridization}), only the $\pi$ orbitals of the {\it A} sublattice, $a_{i\sigma}$, appear since we have assumed the missing carbon atom to be on the {\it B} sublattice. 
Finally, the effective Hamiltonian of graphene with the point defect becomes $H=H_{\rm{gra}}+H_{\rm{def}}+H_{\rm{hyb}}$. 

We apply the numerical renormalization group (NRG) method to this effective Hamiltonian to shed light on the electronic state at low temperatures. 
In the usual Kondo problems, a magnetic impurity interacts with conduction electrons at the same site. 
However, in the present model, the $sp^2$ orbital at the defect interacts with the $\pi$ orbitals of sites 2 and 3. 
Such a geometry is treated in the two-impurity Kondo problem~\cite{Jones1987}. 
We follow the method developed for the two-impurity Kondo problem~\cite{Jones1987,Affleck1995}. 

By diagonalizing in ${\bm k}$ space and retaining low-energy states of the conduction electrons, the Hamiltonian of the conduction electrons (\ref{eq_graphene_hamiltonian}) is expressed as~\cite{Zhu2010,CastroNeto2009}
\begin{eqnarray}
	H_{\rm{gra}}=\sum_{\sigma s=\pm,\tau=1,2}\int_{|{\bm k}|<k_c}\mathrm{d}\bm{ k}\left[\varepsilon_\tau({\bm k})-\mu\right]c^\dagger_{\bm{k}\sigma s\tau}c_{\bm{k}\sigma s\tau},\label{eq_effective_pristine_hamiltonian}
\end{eqnarray}
with the linear dispersion $\varepsilon_\tau({\bm k})=(-1)^\tau v_F|\bm{k}|$, 
where $v_F$ is Fermi velocity, and $s (=\pm)$ and $\tau (=1,2)$ are valley and band indices, respectively. 
Here, the Dirac points are located at ${\bm K}_\pm=(\pm4\pi/3a,0)$ in the Brillouin zone corresponding to the two valleys, with $a$ being the lattice constant, and $\bm{k}$ is measured from $\bm{K}_\pm$ for $s=\pm$. 
The region of the $\bm{k}$ integral is limited to $|\bm{k}|<k_{\rm{c}}$, where $k_{\rm{c}}$ is a cut-off wave number. 
The annihilation operator $a_{i\sigma}$ for the $A$ sublattice is expanded in $c_{{\bm k}\sigma s\tau}$ as 
\begin{eqnarray}
	a_{i\sigma}=\sqrt{\frac{\Omega_{\rm cell}}{2(2\pi)^2}}\sum_{s\tau}\int\mathrm{d}{\bm k}\mathrm{e}^{\mathrm{i}({\bm k}+{\bm K}_s)\cdot{\bm r_i}}c_{{\bm k}\sigma s\tau},
 \end{eqnarray}
 where $\Omega_{\rm cell}$ is the area of the unit cell of graphene. 

In order to express the hybridization term (\ref{eq_hybridization}) in a one-dimensional representation, we introduce the following operator~\cite{Affleck1995}: 
\begin{eqnarray}
	a_{\varepsilon i\sigma}=\sqrt{\frac{\Omega_{\rm cell}}{2(2\pi)^2}}\sum_{s\tau}\int\mathrm{d}{\bm k}\delta(\varepsilon_\tau({\bm k})-\varepsilon)\mathrm{e}^{\mathrm{i}({\bm k}+{\bm K}_s)\cdot{\bm r}_i}c_{{\bm k}\sigma s\tau},\label{eq_energy_rep_site_ope}
\end{eqnarray}
where $\delta(x)$ is the Dirac delta function, and $i=2$ or $3$, ${\bm r}_2={\bm R}/2$, and ${\bm r}_3=-{\bm R}/2$, with ${\bm R}=(a,0)$. 
According to ref.~\citen{Affleck1995} the following orthonormal states are obtained from $a_{\varepsilon2\sigma}$ and $a_{\varepsilon3\sigma}$: 
\begin{eqnarray}
	c_{\varepsilon\sigma e,o}=\frac{1}{N^{1/2}_{e,o}(\varepsilon)}(a_{\varepsilon2\sigma}\pm a_{\varepsilon3\sigma}),\hspace{1em}(\text{even, odd})\label{eq_even_ope}
\end{eqnarray}
where $N_e(\varepsilon)$ and $N_o(\varepsilon)$ are given by
\begin{eqnarray}
	N_{e,o}(\varepsilon)=\frac{\Omega_{\rm cell}}{\pi v^2_F}|\varepsilon|\left[1\mp\frac{1}{2}J_0\left(\frac{|\varepsilon|}{v_F}a\right)\right]. 
\end{eqnarray}
Here, $J_0(x)$ is the zeroth Bessel function. 
Using these operators, the hybridization term (\ref{eq_hybridization}) can be written as 
\begin{eqnarray}
	H_{\rm hyb}=V\sum_{\sigma}\int^D_{-D}\mathrm{d}\varepsilon N^{1/2}_e(\varepsilon)c^\dagger_{\varepsilon\sigma e}d_{\sigma}+\mathrm{h. c. },
\end{eqnarray}
where $D=v_Fk_c$ is a cut-off energy. 
Note that only one channel of the conduction electron ($c_{\varepsilon\sigma e}$) hybridizes with the $sp^2$ orbital, despite there being four species of conduction electrons in graphene ($s$ and  $\tau$). 
This is because the $sp^2$ orbital hybridizes equally with the $\pi$ orbitals at sites $2$ and $3$, as shown in Fig. \ref{fig_distortion}. 
(When the hybridization is not equal, both even and odd states are coupled to the $sp^2$ orbital.) 
Since the dependence of $J_0(|\varepsilon|a/v_F)$ on small $|\varepsilon|a/v_F$ is weaker than $|\varepsilon|$, we approximate $J_0(|\varepsilon|a/v_F)$ as $J_0(0)(=1)$. 
Assuming that $\left(\Omega_{\rm cell}/2\pi v^2_F\right)^{1/2}\simeq1/D$, we obtain
\begin{eqnarray}
	H_{\rm{hyb}}=\frac{V}{D}\sum_\sigma\int^{D}_{-D}\mathrm{d}\varepsilon|\varepsilon|^{1/2}\left(c^\dagger_{\varepsilon\sigma e}d_\sigma+\rm{h.c.}\right).\label{eq_energy_rep_hyb_hamiltonian}
\end{eqnarray}
Using eqs. (\ref{eq_effective_pristine_hamiltonian}), (\ref{eq_energy_rep_site_ope}), and (\ref{eq_even_ope}), it can be shown that $c_{\varepsilon\sigma e}$ satisfies the commutation relation $\left[H_{\rm gra},c^\dagger_{\varepsilon\sigma e}\right]=(\varepsilon-\mu)c^\dagger_{\varepsilon\sigma e}$. 
Thus, we assume that the conduction electron term (\ref{eq_effective_pristine_hamiltonian}) can be written as
\begin{eqnarray}
	H_{\rm{gra}}=\sum_{\sigma}\int^{D}_{-D}\mathrm{d}\varepsilon\left(\varepsilon-\mu\right)c^\dagger_{\varepsilon\sigma e}c_{\varepsilon\sigma e}.\label{eq_energy_rep_pristine_hamiltonian}
\end{eqnarray}
Here, we have neglected the other channels of conduction electrons~\cite{Jones1987, Affleck1995}. 

This model of eqs. (\ref{eq_defect_hamiltonian}), (\ref{eq_energy_rep_hyb_hamiltonian}), and (\ref{eq_energy_rep_pristine_hamiltonian}) is known as the single-channel pseudogap Anderson model~\cite{Gonzalez-Buxton1998} with a finite chemical potential, where the density of states (DOS) of the conduction electron is proportional to $|\varepsilon|$. 
The Kondo problem in such pseudogap systems 
has been studied in detail~\cite{Gonzalez-Buxton1998} for the case with $\mu=0$. 
The main feature in this case is as follows. 
In the particle-hole symmetric ($2\epsilon_{ sp^2}+U=0$) case, no Kondo screening occurs, i.e., the localized moment is not screened by conduction electrons since the DOS vanishes at $\mu=0$. 
On the other hand, in the particle-hole asymmetric ($2\epsilon_{ sp^2}+U\neq0$) case, the localized moment is screened when the coupling constant $V$ is larger than a critical value $V_c$. 
In real materials, the system is expected to be particle-hole asymmetric. 
In ref.~\citen{Gonzalez-Buxton1998}, the particle-hole symmetric case and the $U=\infty$ case were mainly treated, and the case of realistic particle-hole asymmetry has not been studied in detail. 
Hence, we focus on the particle-hole asymmetric case with realistic parameters. 
We also study the chemical potential dependence, which previously has not been studied in detail. 
We apply the NRG method for the model of eqs. (\ref{eq_defect_hamiltonian}), (\ref{eq_energy_rep_hyb_hamiltonian}), and (\ref{eq_energy_rep_pristine_hamiltonian}), following the formalism of ref.~\citen{Gonzalez-Buxton1998}.
The NRG method can include the effects of the pseudogap DOS and a finite chemical potential.

\begin{figure}
	\centering
		\includegraphics[width=6cm]{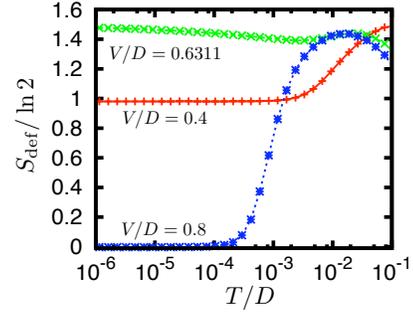}
	\caption{(Color online) Entropy of the electron on the $sp^2$ orbital, $S_{\rm def}$, at chemical potential $\mu=0$ as a function of temperature $T/D$ for the hybridization $V/D=0.4$, $0.6311$, and $0.8$. 
	Here, Boltzmann constant $k_B=1$. }
	\label{fig_entropy_muzero}
\end{figure}
\begin{figure}
	\centering
		\subfigure[$V/D=0.4$]{\includegraphics[width=5cm]{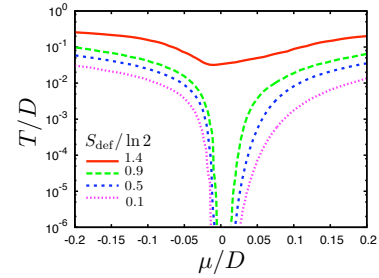}
			\label{fig_entropy_v04}}
		\subfigure[$V/D=0.6311$]{\includegraphics[width=5cm]{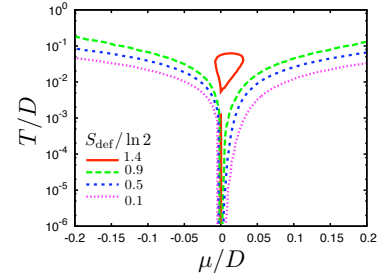}
			\label{fig_entropy_v063}}
		\subfigure[$V/D=0.8$]{\includegraphics[width=5cm]{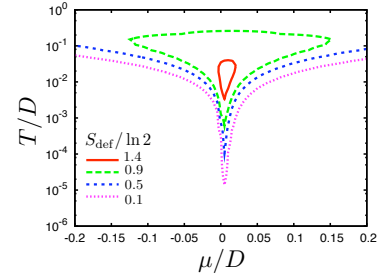}
			\label{fig_entropy_v08}}
	\caption{(Color online) Contour plots of $S_{\rm def}/\ln2$ as a function of chemical potential $\mu/D$ and temperature $T/D$. }
	\label{fig_entropy}
\end{figure}
In order to clarify the low-temperature states of this model in the parameter space of $(V$, $\mu)$, 
we calculate the temperature dependence of the entropy of the electron on the $sp^2$ orbital, $S_{\rm def}$, by the NRG method~\cite{Bulla2008}. 
The bandwidth of graphene is about $D\simeq8$ ${\rm eV}$~\cite{CastroNeto2009}. 
We assume that $\epsilon_{sp^2}\simeq-1$ ${\rm eV}$ and $U\simeq3$ ${\rm eV}$~\cite{Yazyev2007}.
Thus we use the parameters $\epsilon_{sp^2}/D=-0.125$, and $U/D=0.375$, in the following.  
First, we investigate the $\mu/D=0$ case. 
Figure \ref{fig_entropy_muzero} shows $S_{\rm def}$ as a function of temperature for three values of hybridization $V$ ($V/D=0.4$, $0.6311$, and $0.8$).

We find that the critical value of hybridization is $V_c=0.6311D$. 
At $V=V_c$, $S_{\rm def}$ becomes $\ln3$, which is known as the valence-fluctuation fixed point of the pseudogap system~\cite{Gonzalez-Buxton1998}. 
The Boltzmann constant is set to be $k_B=1$. 
For $V<V_c$, the entropy remains $\ln2$ at the lowest temperature ($\sim10^{-6}D$), which indicates an unscreened localized moment. 
For $V>V_c$, $S_{\rm def}$ goes to zero, indicating that the localized moment is screened. 
Despite there being no conduction electron DOS at $\mu=0$, the Kondo screening occurs owing to the particle-hole asymmetry. 

Figure \ref{fig_entropy} shows $S_{\rm def}$ as a function of chemical potential $\mu/D$ and temperature $T/D$ for three values of hybridization $V$ ($V/D=0.4$, $0.6311$, and $0.8$). 
In Figs. \ref{fig_entropy_v04} and \ref{fig_entropy_v063}, we can see that $S_{\rm def}$ goes to zero at low temperature when $\mu\neq0$. 
For $V>V_c$ [Fig. \ref{fig_entropy_v08}], $S_{\rm def}$ goes to zero at low temperatures for all values of $\mu$. 

\begin{figure}
	\centering
		\includegraphics[width=6cm]{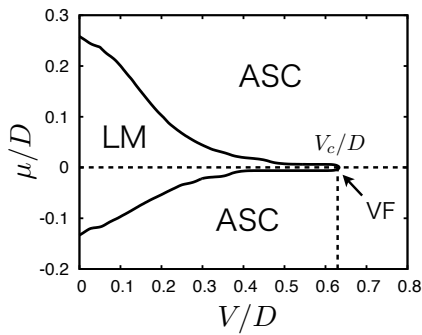}
	\caption{Local-moment (LM) region and asymmetric strong-coupling (ASC) region in the low-temperature hybridization-chemical potential diagram. 
	The valence-fluctuation (VF) fixed point is located at $V_c/D=0.6311, \mu/D=0$. }
	\label{fig_border}
\end{figure}
Combining these data, we plot the low-temperature $V$-$\mu$ phase diagram in Fig. \ref{fig_border}. 
The whole region is divided into the local-moment (LM) region and the asymmetric strong-coupling (ASC) region. 
The valence-fluctuation (VF) fixed point is located at $V=V_c$, $\mu=0$. 

In the experiment~\cite{Chen2011}, it is observed that the Kondo temperature depends on the gate voltage. 
In particular, around the charge-neutrality point ($\mu=0$), the Kondo temperature has a finite  value, and it shows an almost symmetric dependence for the positive and negative gate voltages. 

\begin{figure}
	\centering
		\includegraphics[width=6cm]{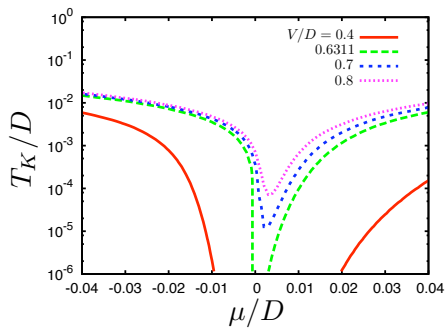}
	\caption{(Color online) Kondo temperature as a function of chemical potential $\mu/D$ for three values of hybridization $V/D$. Here, $\epsilon_{sp^2}/D=-0.125$ and $U/D=0.375$.}
	\label{fig_kondo_temp}
\end{figure}
In order to compare the NRG result with those of the experiment, we plot the Kondo temperature as a function of $\mu/D$ for four values of hybridization $V/D$ in Fig. \ref{fig_kondo_temp}. 
Here, the Kondo temperature, $T_K$, is defined by the relation $S_{\rm def}(T_K)/\ln2=1/2$. 
In the experimental situation, by applying the gate voltage, the chemical potential $\mu$ can be varied in the range of $|\mu|\lesssim0.2$ ${\rm eV}(\simeq0.025D)$. 
As shown in Fig. \ref{fig_kondo_temp}, $T_K\ll10^{-6}D$ for $V\le V_c$ at $\mu=0$. 
However, $T_K$ is finite for $V>V_c$ even at $\mu=0$. 
This behavior reproduces the experimental behavior. 
We find the almost symmetric dependence on the chemical potential. 
Note that this feature is not obtained in the $U=\infty$ case~\cite{Vojta2010}, where $T_K$ becomes totally asymmetric. 

In this study, we have considered only the most stable Jahn-Teller distorted structure, i. e., $h\neq h'$. 
However, the threefold symmetric structure may be realized under certain conditions. 
In such a case, there is a degeneracy of the molecular orbitals at the defect site. 
It can cause a more exotic Kondo effect, such as the three-impurity Kondo effect~\cite{Ingersent2005}. 
This remains as a future problem. 
Also, the dynamical effects of the lattice degrees of freedom at the defect site have been neglected. 
Actually, the dynamical switching of the covalent bond (dynamical Jahn-Teller effect) in the point defect has been discussed~\cite{El-Barbary2003,Amara2007}. 
We speculate that this effect enhances the electron scattering at the defect site and hence enhances the Kondo effect~\cite{Hotta2006}. 

In conclusion, a point defect in graphene has been investigated as a cause of the Kondo effect. 
Taking account of the electronic state and the lattice structure of the defect, an effective model has been developed. 
This model is transformed into the single-channel pseudogap Anderson model with a finite chemical potential. 
By numerical renormalization group calculations, the low-temperature phase diagram of this model is determined in the whole parameter regions of hybridization and chemical potential. 
It is shown that the experimentally observed gate-voltage dependence of the Kondo temperature can be reproduced in this framework.

\begin{acknowledgements}
We are grateful to K. Miyake for stimulating conversations and discussions. 
One of the authors (T. K.) is grateful to T. Kariyado for technical advice. 
T. K. is supported by the Global COE program \lq\lq the Physical Sciences Frontier" of the Ministry of Education, Culture, Sports, Science and Technology, Japan. 
\end{acknowledgements}


\begin{thebibliography}{50}
	\bibitem{Sengupta2008}
		K. Sengupta and G. Baskaran: Phys. Rev. B {\bf 77} (2008) 045417. 
	\bibitem{Vojta2010}
		M. Vojta, L. Fritz, and R. Bulla: Europhys. Lett. {\bf 90} (2010) 27006. 
	\bibitem{Zhu2010}
		Z.-G. Zhu, K.-H. Ding, and J. Berakdar: Europhys. Lett. {\bf 90} (2010) 67001. 
	\bibitem{Haase2011}
		P. Haase, S. Fuchs, T. Pruschke, H. Ochoa, and F. Guinea: Phys. Rev. B {\bf 83} (2011) 241408. 
	\bibitem{Kotov2010}
		V. N. Kotov, B. Uchoa, V. M. Pereira, F. Guinea, and A. H. Castro Neto: arXiv:1012.3484v2. 
	\bibitem{CastroNeto2009}
		A. H. Castro Neto, F. Guinea, N. M. R. Peres, K. S. Novoselov, and A. K. Geim: Rev. Mod. Phys. {\bf 81} (2009) 109; 
		D. S. L. Abergel, V. Apalkov, J. Berashevich, K. Ziegler, and
T. Chakraborty: Adv. Phys. {\bf 59} (2010) 261. 
	\bibitem{Chen2011}
		J.-H. Chen, L. Li, W. G. Cullen, E. D. Williams, and M. S. Fuhrer: Nat. Phys. {\bf 7} (2011) 535. 
	\bibitem{Nair2012}
		R. R. Nair, M. Sepioni, I-L. Tsai, O. Lehtinen, J. Keinonen, A. V. Krasheninnikov, T. Thomson, A. K. Geim, and I. V. Grigorieva: Nat. Phys. {\bf 8} (2012) 199. 
	\bibitem{Watkins1963}
		G. D. Watkins: {\it Proc. Int. Conf. Crystal Lattice Defects}, J. Phys. Soc. Jpn. {\bf 18} (1963) Suppl. II, p. 22; 
		M. Lannoo and J. Bourgoin: {\it Point Defects in Semiconductors I, II} (Springer-Verlag, Berlin, 1981, 1983). 
	\bibitem{Ugeda2010}
		M. M. Ugeda, I. Brihuega, F. Guinea, and J. M. G\'omez-Rodr\'iguez: Phys. Rev. Lett. {\bf 104} (2010) 096804. 
	\bibitem{Lehtinen2004}
		P. O. Lehtinen, A. S. Foster, Y. Ma, A. V. Krasheninnikov, and R. M. Nieminen: Phys. Rev. Lett. {\bf 93} (2004) 187202. 
	\bibitem{Yazyev2007}
		O. V. Yazyev and L. Helm: Phys. Rev. B {\bf 75} (2007) 125408. 
	\bibitem{Wilson1975}
		K. G. Wilson: Rev. Mod. Phys. { \bf 47} (1975) 773.
	\bibitem{Meyer2008}
		J. C. Meyer, C. Kisielowski, R. Erni, M. D. Rossell, M. F. Crommie, and A. Zettl: Nano Lett. {\bf 8} (2008) 3582. 
	\bibitem{El-Barbary2003}
		A. A. El-Barbary, R. H. Telling, C. P. Ewels, M. I. Heggie, and P. R. Briddon: Phys. Rev. B {\bf 68} (2003) 144107. 
	\bibitem{Jones1987}
	B. A. Jones and C. M. Varma: Phys. Rev. Lett. {\bf 58} (1987) 843. 
	\bibitem{Affleck1995}
		I. Affleck, A. W. W. Ludwig, and B. A. Jones: Phys. Rev. B {\bf 52} (1995) 9528. 
	\bibitem{Gonzalez-Buxton1998}
		C. Gonzalez-Buxton and K. Ingersent: Phys. Rev. B {\bf 57} (1998) 14254. 
	\bibitem{Bulla2008}
		R. Bulla, T. A. Costi, and T. Pruschke: Rev. Mod. Phys. {\bf 80} (2008) 395. 
	\bibitem{Ingersent2005}
		K. Ingersent, A. W. W. Ludwig, and I. Affleck: Phys. Rev. Lett. {\bf 95} (2005) 257204. 
	\bibitem{Amara2007}
		H. Amara, S. Latil, V. Meunier, P. Lambin, and J.-C. Charlier: Phys. Rev. B {\bf 76} (2007) 115423. 
	\bibitem{Hotta2006}
		T. Hotta: Phys. Rev. Lett. {\bf 96} (2006) 197201; 
		T. Hotta: J. Phys. Soc. Jpn. {\bf 76} (2007) 084702. 
\end{thebibliography}
\end{document}